\documentclass[twocolumn,english,aps,prl,floatfix,showpacs]{revtex4}
\usepackage[T1]{fontenc}
\usepackage[latin1]{inputenc}

\makeatletter

\usepackage{babel}

\usepackage{babel}
\makeatother
\begin{document}

\title{Comment on 'Coherent-state path integrals in the continuum'}

\author{ Evgenii Kochetov}

\affiliation{ Bogoliubov Laboratory of Theoretical Physics, Joint
Institute for Nuclear Research, 141980 Dubna, Russia}


\maketitle

Coherent states (CSs) associated with a Lie algebra of a group $G$ appear as points of orbits of a unitary irreducible representation of $G$ in an abstract Hilbert space.
The associated CS path integral provides an efficient way to (quasiclassically) quantize  such orbits --  $G$-homogeneous phase spaces --with $G$ acting as a group of canonical transformations.
The most important examples are provided by the $SU(2)$ group  generated by the conventional spin operators, $\vec S$, and the Heisenberg-Weyl (HW) group
generated by $I, b, b^{\dagger}, b^{\dagger}b$ with $I$ being the identity and $b$ being a bosonic operator.  The related continuous CS path-integrals  work well
provided the dynamics is governed by a linear combination (with time-dependent coefficients) of the group generators.
However the continuous CS path integral seems to fail to produce a correct result in case a square of the  Cartan operator is involved.
For example, this is the case for the Hamiltonian $H$ that comprises the $A^2$ term, with $A=S_z\, (s>1/2)$ or $A= b^{\dagger}b$ representing the
Cartan elements of the $su(2)$ algebra  and the HW algebra, respectively.
In such case there is an "anomaly" in the definition of the CS path integral in the continuum which was first reported in \cite{wilson}.

The authors of the recent paper \cite{kordas}  claim to have established a time continuous formulation of path integration in the CS basis free from the mentioned inconsistency. Since a few recent investigations consider the approach advocated in \cite{kordas} as providing a "remedy" for the
CS path-integral anomaly (see, e.g.,  \cite{masch}), I feel it is appropriate to show that the "recipe" proposed in \cite{kordas} does not eliminate the anomaly. The origin of the anomaly detected in \cite{wilson} still remains unclear.

Let us for concreteness consider a single-site Bose-Hubbard (BH) model,
$ H_{BH} =\frac{U}{2}n(n-1)$, with $n=b^{\dagger}b.$ The action in the continuous CS path integral for the partition function reads
$ S_{cov} =\langle z|-\partial_{\tau} - H|z\rangle,$
where $|z\rangle$ is the coherent state associated with the HW algebra. The subscript "cov" indicates that a covariant (Wick) symbol
of the Hamiltonian is involved, $H_{cov}=\langle z|H|z\rangle.$  This prescription amounts to
the fact that the discontinuity of the correlator $\langle b(\tau) \bar b(\tau')\rangle $, at equal time arguments,
should be dealt with according to the rules following from the defining time-lattice approximation of the
corresponding coherent-state path integral.\cite{negele} Namely, the Green function  at coinciding time arguments
should be taken in the form $G(\tau,\tau+\epsilon), \epsilon\to 0^+$. This is a consequence of the fact that the bar field
$\bar z(\tau_{k+1})$ in the time-lattice path integral action
comes from the next time slice compared to the field $z(\tau_k)$. This important observation
explicitly demonstrates that a continuum form of a coherent-state path integral cannot be considered independently of
the underlying time-lattice approximation (see, e.g., \cite{negele}).

Wilson and Galitski explicitly demonstrated that the anomaly in the CS path integral disappears, provided a covariant symbol of the operator product is replaced by a product of the symbols.\cite{wilson} However $\langle z | n^2|z\rangle \neq  \langle z | n|z\rangle ^2.$ The similar observation holds for the $su(2)$ CS path integral as well: it is free from the anomaly if $(S_z^2)_{cov}\to (S_z)^2_{cov}.$

In principle other symbols besides the covariant ones  can be used in  the continuous path-integral representation of the partition function. This   amounts to employing  different quantization schemes.
In paper \cite{kordas}, the Weyl or symmetric quantization is invoked to deal with the CS path integral for the BH model.
In the continuum limit, the underlying time-lattice approximation manifests itself through
a Green function symmetrized at the coinciding time arguments,
$G(\tau,\tau)=\frac{1}{2}(G(\tau,\tau+\epsilon)+G(\tau+\epsilon,\tau))),\, \epsilon \to 0^+.$
It is precisely this prescription that gives a concrete meaning to the formal CS path integrals used in \cite{kordas}.
It is related to a certain regularization of the determinant that emerges upon integrating out the $z$-field.\cite{elitzur}

To make this point explicit let us consider a continuum bosonic CS path integral with periodic boundary conditions,
\begin{equation}
I_{\mu}=\int D\bar zDz \exp{(-\int_0^{\beta}\bar z(\partial_{\tau} +\mu)zd\tau)}.
\label{1}
\end{equation}
At first glance a specific underlying time-lattice form of this integral is inessential since it does not seem to affect (\ref{1}).
We now show that this is not the case, however.
Formally,
\begin{equation}
I_{\mu}/I_{\mu_0}=\exp{(-\int_{\mu_0}^{\mu} d\mu\, tr G)},
\label{2}\end{equation}
where
$G:=(\partial_{\tau} +\mu)^{-1}, \, tr G=\int_0^{\beta}d\tau G(\tau,\tau)=\beta G(0),$
$G(\tau_1,\tau_2)= G(\tau_1-\tau_2)=[\frac{1}{e^{\beta\mu}-1}+\theta(\tau_1-\tau_2)]e^{-\mu(\tau_1-\tau_2)}.$
We thus see that to compute (\ref{1}) we need to know the Green function $G$ at coinciding time arguments. However the theta function that
enters $G$ is discontinuous at coinciding arguments. To fix this discontinuity we need some extra information that is implicitly encoded
in the underlying time-lattice approximation. For example, in the covariant quantization scheme, the corresponding time-lattice representation implies that
$G(0)$  should be taken as $G(0^-)$ so that $I_{\mu}/I_{\mu_0}= (1-e^{-\beta\mu_0})/(1-e^{-\beta\mu}).$
The contravariant quantization implies $G(0)\to G(0^+)$, so that $I_{\mu}/I_{\mu_0}= (1-e^{-\beta\mu_0})e^{-\beta\mu}/(1-e^{-\beta\mu})e^{-\beta\mu_0}.$
Within the symmetric (Weyl) quantization
used by the authors, $G(0)\to (G(0^+)+G(0^{-}))/2 $ which yields $ I_{\mu}/I_{\mu_0}= (1-e^{-\beta\mu_0})e^{-\beta\mu/2}/(1-e^{-\beta\mu})e^{-\beta\mu_0/2}$ as it should be. We thus see that as long as the underlying time-lattice structure is properly taken into consideration no ambiguity in the continuum CS path integral arises.

In real $(\hat p,\hat q)$ coordinates, the BH Hamiltonian is given by their Eq.(15). The authors then arrive at the path integral with the classical action given
by Eq.(19). In doing so they simply replace the operators $\hat p$ and $\hat q$ by the corresponding $c$-numbers.
This amounts to
replacing $((\hat p^2+\hat q^2)^2)_W$ by the product of the symbols   $(\hat p^2+\hat q^2)_W(\hat p^2+\hat q^2)_W$, which is an incorrect procedure, however.  Actually a symbol of a product of operators differs from a product of the symbols.
The latter quantity does not even represent a symbol of a certain operator at all.
If the authors computed the Weyl symbol of a square of the operators correctly the anomaly would immediately pop up again.
Let us show this explicitly.
The covariant and Weyl symbols are, in a flat phase space,  related to each other through the equation
$H_W(\bar z,z)=e^{-\frac{1}{2}\Delta} H_{cov}(\bar z, z),$ where $\Delta=\frac{\partial^2}{\partial\bar z\partial z}$ is the Laplace-Beltrami operator. \cite{ber}  Computing this,  yields (here we put $\mu=0$ and $U=1$)
$H^{BH}_W=-|z|^2+1/2|z|^4+1/4,$ which disagrees with their  Eq.(19) by a constant. As a result we observe the anomaly in the path integral.
Let us now use instead the  "linearized"   BH "Weyl symbol":
$\tilde H_{W}^{BH}=\frac{1}{2}(n_Wn_W -n_W)$.
Since $n_W=|z|^2-1/2$
we arrive exactly, in this way,  at
Eq.(19).  The anomaly has thus been  eliminated making use of this unjustified approximation.

In the paper \cite{kordas},  the function  $\tilde H_W^{BH}$ is denoted by $H_{BH}^F$. It is claimed to represent the classical
Hamiltonian that enters the path-integral action.
However one should keep in mind that
an action that enters the CS path integral is necessarily represented by a symbol of a Hamiltonian rather than a classical Hamiltonian function (see, e.g.
\cite{berezin}). Just replacing the $(\hat p, \hat q)$ operators in Eq.(15)  by the corresponding $c$-numbers is not an appropriate procedure.
If the path-integral quantization could have been defined through such a procedure (that ignores the $\hat p,\hat q$ ordering altogether), this would have established  a one-to-one correspondence between classical mechanics and quantum mechanics making the latter an inessential ingredient.
Quantum mechanics can however be thought of as a noncommutative extension of the  Poisson algebra of classical observables and that extension is realized through the appropriate choice of operator symbols.

In summary,  the Weyl CS path-integral quantization is not free from the anomaly observed in \cite{wilson}. The attempts exposed in  \cite{kordas} to prove the contrary are founded on the unjustified approximations in dealing with the CS path integral. Their "recipe" given by Eq.(20)
to associate a quantum Hamiltonian with a classical function  only works provided $H^{F}=H_W$ which is the case for noninteracting quadratic Hamiltonians. In our case, however, $H^F=\tilde H_W \neq H_{W}$ and the route (20) cannot be justified.

\end{document}